# Multiphoton super-resolution imaging via virtual structured illumination

Sumin Lim[1,2], Sungsam Kang[1,2], Jin-Hee Hong[1,2], Youngho Jin[1], Kalpak Gupta[1,2], Moonseok Kim[3], Suhyun Kim[4], Wonshik Choi[1,2*], Seokchan Yoon[5*]

[1]*Center for Molecular Spectroscopy and Dynamics, Institute for Basic Science, Seoul 02841, Korea*

[2]*Department of Physics, Korea University, Seoul 02855, Korea*

[3]*Department of Medical Life Sciences, College of Medicine, The Catholic University of Korea, Seoul, 06591, Korea*

[4]*Department of Biomedical Sciences, Korea University, Ansan, Republic of Korea*

[5]*School of Biomedical Convergence Engineering, Pusan National University, Yangsan 50612, Korea*

[*]*Corresponding author(s). E-mail(s): wonshik@korea.ac.kr; sc.yoon@pusan.ac.kr*

## Abstract

Imaging in thick biological tissues is often degraded by sample-induced aberrations, which reduce image quality and resolution, particularly in super-resolution techniques. While hardware-based adaptive optics, which correct aberrations using wavefront shaping devices, provide an effective solution, their complexity and cost limit accessibility. Computational methods offer simpler alternatives but struggle with complex aberrations due to the incoherent nature of fluorescence. Here, we present a deep-tissue super-resolution imaging framework that addresses these challenges with minimal hardware modification. By replacing the photodetector in a standard laser-scanning microscope with a camera, we measure an incoherent response matrix (IRM). A dual deconvolution algorithm is developed to decompose the IRM into excitation and emission optical transfer functions and the object spectrum. The proposed method simultaneously corrects excitation and emission point-spread functions (PSFs), achieving a resolution of $\lambda/4$, comparable to structured illumination microscopy. Unlike existing computational methods that rely on vector decomposition of a single convoluted PSF, our matrix-based approach enhances image reconstruction, particularly for high spatial frequency components, enabling super-resolution even in the presence of complex aberrations. We validated this framework with two-photon super-resolution imaging, achieving a lateral resolution of 130 nm at a depth of 180 μm in thick mouse brain tissue.

## Introduction

Fluorescence imaging has become an indispensable tool in biology and life sciences due to its molecular specificity and high contrast. One major challenge in fluorescence imaging has been overcoming the diffraction limit of resolution, which has been addressed by several super-resolution imaging techniques developed over the past decades. Structured illumination microscopy (SIM)[1] achieves super-resolution by illuminating the target with structured light patterns and expanding the bandwidth via the synthesis of spatial frequency information. Techniques like image scanning microscopy[2, 3] employ pixel reassignments[4], which can be viewed as the point-illumination version of SIM[5]. Stimulated emission depletion (STED)[6] microscopy employs nonlinear point-spread-function (PSF) engineering to achieve foci below the diffraction limit, while single-molecule localization microscopy (SMLM)[7, 8] relies on the localization of individual fluorophores through stochastic activation or switching. Additionally, super-resolution fluctuation imaging[9] achieves enhanced resolution by analyzing high-order temporal correlations of fluorescence signals.

A major focus in fluorescence imaging is enhancing imaging depth to visualize structures deep within biological tissues[10]. However, sample-induced aberrations distort the shape and peak intensity of the PSFs, thereby degrading resolution and depth[11]. For example, two-photon fluorescence microscopy (2PFM) suffers from reduced signal and resolution due to distortion of excitation PSFs caused by tissue aberrations[12, 13]. In SIM, illumination and emission patterns experience independent distortions, arising from the difference in wavelengths between the illumination light and emitted fluorescence, leading to a loss of super-resolution capability[14, 15]. Aberrations are even more detrimental to STED microscopy and SMLM, as their nonlinearity and weak single-molecule emission signals, respectively, make them more susceptible to PSF perturbation[16,17].

Sample-induced aberrations in fluorescence imaging have mostly been addressed using hardware-based adaptive optics (AO)[18-20], which employ wavefront shaping devices such as deformable mirrors and liquid-crystal spatial light modulators. To determine the wavefront corrections to be applied to the wavefront shaping devices, either wavefront sensing[20-22] or sensorless methods[11, 19] have been developed. Wavefront sensing methods use devices such as Shack-Hartmann sensors[11, 19, 23] or interferometric systems[24], while sensorless methods iteratively optimize wavefront shaping devices based on image brightness or sharpness metrics[11]. In both cases, hardware-based AO approaches increase system complexity and require additional procedures before imaging, making them less accessible for biologists.

To avoid the need for additional hardware components, computational correction of aberrations from acquired fluorescence imaging data is ideal. However, fluorescence imaging lacks wavefront information due to the intrinsically incoherent nature of fluorescence emission, making it difficult to directly obtain wavefront distortions as in coherent imaging. Instead, computational AO reconstruction methods, such as machine learning-based AO method[25, 26] or deconvolution methods[27-33] are commonly used. The core idea is to estimate a PSF and an object image that best fit the measured fluorescence image. However, these cannot independently handle the excitation PSF and the emission PSF; instead, they address only a single PSF, which is the convolution of the two. This convoluted PSF typically has a narrower spectral bandwidth than the individual excitation and emission PSFs. As a result, high-frequency information is often lost during acquisition, making it computationally difficult, or even impossible, to fully recover. Furthermore, conventional deconvolution is a vector decomposition that is intrinsically underdetermined and requires prior knowledge of the PSF. Due to these limitations, existing conventional image reconstruction methods are effective under conditions of mild aberration or simple PSF perturbations.

We propose a novel computational AO framework for deep-tissue super-resolution imaging. Our approach replaces the integral detector used in conventional laser-scanning microscopy (e.g., photomultiplier tubes) with a camera and measures an incoherent response matrix (IRM). We developed a dual deconvolution algorithm to decompose the measured IRM into the excitation and emission optical transfer functions (OTFs) and the object spectrum. In doing so, the proposed method achieves a resolution of $\lambda/4$, comparable to structured illumination microscopy. In contrast to traditional computational methods, which rely on vector decomposition based on a single convoluted PSF, our dual deconvolution approach performs matrix decomposition, simultaneously correcting both excitation and emission PSFs. This substantially enhances image reconstruction, particularly in high spatial frequency components, restoring super-resolution in complex aberrations that existing methods cannot handle. We validated this framework in two-photon fluorescence microscopy, achieving multi-photon super-resolution imaging by visualizing dendritic spines in thick mouse brain tissue, with a spatial resolution of 130 nm at a depth of 180 μm.

## Results

### Incoherent Response Matrix Formalism for SIM

The proposed method begins with an imaging configuration where the sample is illuminated with a focused beam, and the generated fluorescence signal is measured by an array detector. Consider that a tightly focused excitation laser beam is directed to a position $\mathbf{r}_\mathrm{i}$ in the plane conjugate to the sample plane with spatial coordinate $\mathbf{r}$ where the fluorescent objects are embedded within a scattering medium (Fig. 1a). The excitation laser beam is distorted due to sample-induced aberration, characterized by the excitation point spread function (PSF), $h_\mathrm{ex}(\mathbf{r})$. This distorted beam interacts with the fluorescent targets at the sample plane, which have a fluorophore density distribution $\gamma(\mathbf{r})$. The emitted fluorescence from the targets experiences further aberration, described by the emission PSF, $h_\mathrm{em}(\mathbf{r})$. Consequently, the fluorescence intensity map at the detector plane $\mathbf{r}_\mathrm{d}$ for each point-illumination position $\mathbf{r}_\mathrm{i}$ is described by:

$$f(\mathbf{r}_\mathrm{d},\mathbf{r}_\mathrm{i}) = \int h_\mathrm{em}(\mathbf{r}_\mathrm{d}-\mathbf{r})\gamma(\mathbf{r})h_\mathrm{ex}(\mathbf{r}-\mathbf{r}_\mathrm{i})\,\mathrm{d}\mathbf{r}. \quad (1)$$

We record a set of fluorescence images for different illumination positions and construct the incoherent response matrix (IRM) whose elements are given by $f(\mathbf{r}_\mathrm{d};\mathbf{r}_\mathrm{i})$. Essentially, this IRM contains the interrelation of both excitation and emission PSFs in the spatial domain in the form of convolution.

To make the relation between IRM and PSFs concise, the IRM can be converted from spatial domain to spatial frequency domain. Once the IRM is obtained, we can computationally synthesize a fluorescence image $I_\mathrm{fl}(\mathbf{r}_\mathrm{d})$ for an arbitrary incoherent illumination pattern $I_\mathrm{ill}(\mathbf{r}_\mathrm{i})$: $I_\mathrm{fl}(\mathbf{r}_\mathrm{d}) = \int f(\mathbf{r}_\mathrm{d};\mathbf{r}_\mathrm{i})I_\mathrm{ill}(\mathbf{r}_\mathrm{i})\mathrm{d}\mathbf{r}_\mathrm{i}$. As a special case, we consider sending a sinusoidally modulated illumination $I_\mathrm{ill}(\mathbf{r}_\mathrm{i}) = e^{i\mathbf{k}_\mathrm{i}\cdot\mathbf{r}_\mathrm{i}}$ and obtain wide-field fluorescence image, $I_\mathrm{fl}(\mathbf{r}_\mathrm{d},\mathbf{k}_\mathrm{i})$, for each $\mathbf{k}_\mathrm{i}$ (Fig. 1b). By taking the Fourier transform of each fluorescence image with respect to $\mathbf{r}_\mathrm{d}$, we can obtain the IRM in the spatial frequency domain, termed spectral IRM:

$$F(\mathbf{k}_\mathrm{d},\mathbf{k}_\mathrm{i}) = H_\mathrm{em}(\mathbf{k}_\mathrm{d})H_\mathrm{ex}(\mathbf{k}_\mathrm{i})\Gamma(\mathbf{k}_\mathrm{d},\mathbf{k}_\mathrm{i}). \quad (2)$$

Here, $H_\mathrm{ex}$, $\Gamma$, and $H_\mathrm{em}$ are Fourier transforms of $h_\mathrm{ex}$, $\gamma$, and $h_\mathrm{em}$, respectively. Therefore, $H_\mathrm{ex}$ and $H_\mathrm{em}$ correspond to the excitation and emission OTFs, respectively. In fact, this acquisition of $F(\mathbf{k}_\mathrm{d},\mathbf{k}_\mathrm{i})$ is equivalent to the working principle of SIM[34]. In SIM, raw images are recorded by illuminating sinusoidal intensity patterns $I_\mathrm{ill}(\mathbf{r}_\mathrm{i}) = 1+\cos(\mathbf{k}_\mathrm{i}\cdot\mathbf{r}_\mathrm{i}+\phi)$ for a few spatial frequencies $\mathbf{k}_\mathrm{i}$ with known pattern phases $\phi$ on the sample, and $F(\mathbf{k}_\mathrm{d},\mathbf{k}_\mathrm{i})$ is obtained by the demodulation process with respect to $\phi$.

In case when there are no aberration and scattering, the object spectrum can be obtained by aperture synthesis of the spectral IRM:

$$\Gamma(\Delta\mathbf{k}) = \frac{1}{H_\mathrm{eff}(\Delta\mathbf{k})}\sum_{\mathbf{k}_\mathrm{i}} F(\mathbf{k}_\mathrm{i}+\Delta\mathbf{k},\mathbf{k}_\mathrm{i}), \quad (3)$$

where $\Delta\mathbf{k} = \mathbf{k}_\mathrm{d}-\mathbf{k}_\mathrm{i}$, and $H_\mathrm{eff}(\Delta\mathbf{k})$ is the effective OTF, given by the convolution of the emission and excitation OTFs, $H_\mathrm{eff}(\Delta\mathbf{k}) = \sum_{\mathbf{k}} H_\mathrm{em}(\mathbf{k}+\Delta\mathbf{k})H_\mathrm{ex}(\mathbf{k})$. The bandwidth of the synthesized object spectrum is extended up to $k_\mathrm{c} = 2\alpha(k_\mathrm{em}+k_\mathrm{ex})$, where $\alpha$ is the numerical aperture, and $k_\mathrm{em} = 2\pi/\lambda_\mathrm{em}$ and $k_\mathrm{ex} = 2\pi n/\lambda_\mathrm{ex}$ are the diffraction-limited cut-off spatial frequencies for the emission and excitation wavelengths, $\lambda_\mathrm{em}$ and $\lambda_\mathrm{ex}$, respectively. This framework can describe multi-photon fluorescence imaging by replacing the excitation PSF with $h_{ex} = h_{ex}^n$, where $n$ is the photon excitation order ($n = 2$ for two-photon excitation). In our study, we mainly focused on two-photon fluorescence imaging as it is better suited for deep-tissue imaging. The resulting spatial resolution is $\lambda_{ex}\lambda_{em}/2n\alpha(\lambda_{em} + \lambda_{ex}/n)$. In the case of two-photon excitation with $\lambda_{ex}/2 \approx \lambda_{em}$ and $\alpha = 1$, the spatial resolution approaches $\lambda_{em}/4$, surpassing the diffraction limit.

**Retrieval of Excitation and Emission OTFs from IRM**

The sample-induced aberrations modify the phase part of the OTF, known as the phase transfer function (PTF), which distorts PSF shape. They also attenuate the amplitude part of the OTF, termed the modulation transfer function (MTF), especially at high spatial frequencies. This leads to a reduced bandwidth of the measured object spectrum, thereby compromising the resolving power. In fluorescence imaging, the excitation and emission PSFs are affected independently due to their wavelength differences, especially in multi-photon imaging. Therefore, separately retrieving and correcting $h_{ex}$ and $h_{em}$ is critical for achieving full theoretical bandwidth resolution. To this end, we propose the dual-deconvolution algorithm, which separately identifies $H_{em}$ and $H_{ex}$ from the spectral IRM in Eq. (2). The key principle of dual deconvolution is based on the iterative Wiener filter method[35, 36], solving the Wiener-filter like equation in a matrix form:

$$F(\mathbf{k}_d, \mathbf{k}_i) = H(\mathbf{k}_d, \mathbf{k}_i)\Gamma(\mathbf{k}_d, \mathbf{k}_i), \quad (4)$$

subject to $H(\mathbf{k}_d, \mathbf{k}_i) = H_{em}(\mathbf{k}_d)H_{ex}(\mathbf{k}_i)$ and $\Gamma(\mathbf{k}_d, \mathbf{k}_i) = \Gamma(\Delta\mathbf{k})$. We introduce a matrix Wiener restoration filter $W(\mathbf{k}_d, \mathbf{k}_i)$ to estimate $\Gamma$ from the measured $F$ by $\Gamma(\mathbf{k}_d, \mathbf{k}_i) = W(\mathbf{k}_d, \mathbf{k}_i)F(\mathbf{k}_d, \mathbf{k}_i)$. Then, $H_{em}$ and $H_{ex}$ are estimated from the row and column correlations between $F(\mathbf{k}_d, \mathbf{k}_i)$ and $\Gamma(\mathbf{k}_d, \mathbf{k}_i)$. By executing this process iteratively, our dual-deconvolution algorithm gradually refines the estimated OTFs and $\Gamma$ (see Methods for details).

The conventional blind deconvolution algorithm used in confocal imaging relies on element-wise vector decomposition, whereas our dual-deconvolution algorithm utilizes element-wise matrix decomposition. This distinction has a substantial impact on the recoverable resolution, especially in the presence of complex aberrations. To make this point clear, let us explain the image formation in confocal imaging. A confocal image can be described as $i_{con}(\mathbf{r}_i) = f(\mathbf{r}_d = \mathbf{r}_i, \mathbf{r}_i)$:

$$i_{con}(\mathbf{r}_i) = \int h(\mathbf{r} - \mathbf{r}_i)\gamma(\mathbf{r})\mathrm{d}\mathbf{r}, \quad (4)$$

where $h(\mathbf{r})$ is the effective PSF, given by $h(\mathbf{r}) = h_{em}(-\mathbf{r})h_{ex}(\mathbf{r})$. The spectrum of the confocal image is given by $I_{con}(\Delta\mathbf{k}) = H_{eff}(\Delta\mathbf{k})\Gamma(\Delta\mathbf{k})$. Here, $H_{eff}(\Delta\mathbf{k})$ is the effective OTF, given by $H_{eff}(\Delta\mathbf{k}) = \int H_{em}(\mathbf{k} + \Delta\mathbf{k})H_{ex}(\mathbf{k})\mathrm{d}\mathbf{k}$. To recover the original object image from the acquired image, blind deconvolution such as joint Richardson-Lucy deconvolution[27, 28] is commonly employed, which estimates both the unknown $H_{eff}$ and $\Gamma$ from $I_{con}$. This vector decomposition is generally underdetermined and works only when prior knowledge of the PSF is available. Furthermore, the effective OTF ($H_{eff}$) can experience significant attenuation at high spatial frequencies during image acquisition, as each element of $H_{eff}$ is the result of the summation of the product of two complex-valued functions, $H_{em}$ and $H_{ex}$, through deconvolution. Consequently, frequency components below the noise level may not be properly restored or could be entirely lost. In contrast, our dual-deconvolution algorithm is a matrix decomposition, a well-determined problem. Therefore, it works for arbitrary complex aberrations with no prior knowledge or assumption. Furthermore, the OTF matrix $H(\mathbf{k}_d, \mathbf{k}_i)$ in IRM is simply the product of $H_{em}$ and $H_{ex}$, allowing better preservation of high-frequency content. By leveraging this preservation, our dual-deconvolution algorithm can recover much higher frequency components compared to conventional blind deconvolution algorithms.

**Numerical Validation of Dual Deconvolution**

Figure 2 illustrates the validation of the proposed algorithm with simulated two-photon IRM following our proposed point-scanning geometry in Fig. 1a (see Methods for details of numerical simulation). This simulated data considered an emission wavelength of $\lambda_{\mathrm{em}} = 520$ nm and $\alpha = 1$. For simplicity, we assume that the two-photon excitation wavelength is twice the emission wavelength, i.e., $\lambda_{\mathrm{ex}} = 2\lambda_{\mathrm{em}}$ such that the cut-off spatial frequencies for excitation and emission OTFs are the same. Therefore, the theoretically achievable resolution of a reconstructed two-photon SIM (2PSIM) image is a quarter of the fluorescence wavelength, $\lambda_{\mathrm{em}}/4 = 130$ nm. The procedure to generate two-photon IRM in the presence of excitation and emission aberrations is as follows. We computationally added two independent phase aberrations, $\varphi_{\mathrm{ex}}(\mathbf{k})$ and $\varphi_{\mathrm{em}}(\mathbf{k})$, to excitation and emission pupils, respectively, which were generated by the superposition of Zernike modes up to the order of 30 with various mode coefficients. The absolute squares of the Fourier transform of the complex pupil functions lead to the one-photon excitation PSF and the emission PSF. To simulate two-photon fluorescence imaging, we obtained the two-photon excitation PSF by taking the square of the one-photon excitation PSF. A set of fluorescence intensity maps, i.e., the incoherent response matrix $f(\mathbf{r}_{\mathrm{d}}; \mathbf{r}_{\mathrm{i}})$, was obtained for a given target function, $\gamma(\mathbf{r})$, using Eq. (1). Finally, the spectral IRM $F(\mathbf{k}_{\mathrm{o}}; \mathbf{k}_{\mathrm{i}})$ was constructed by Fourier transforming $f(\mathbf{r}_{\mathrm{d}}; \mathbf{r}_{\mathrm{i}})$.

We first constructed an image equivalent to the conventional 2PFM image (Fig. 2a) by summing all pixel values across each column of the IRM, i.e. $I_{\mathrm{2PFM}}(\mathbf{r}_{\mathrm{i}}) = \sum_{\mathbf{r}_{\mathrm{d}}} f(\mathbf{r}_{\mathrm{d}}; \mathbf{r}_{\mathrm{i}})$. A 2PSIM image (Fig. 2b) was obtained by inverse Fourier transforming the aperture-synthesized object spectrum, $I_{\mathrm{2PSIM}}(\Delta\mathbf{k}) = \sum_{\mathbf{k}_{\mathrm{i}}} F(\Delta\mathbf{k} + \mathbf{k}_{\mathrm{i}}; \mathbf{k}_{\mathrm{i}})$. Both the 2PFM and 2PSIM images were blurred due to aberrations.

Next, we applied our dual-deconvolution algorithm to the simulated spectral IRM to estimate the excitation and emission OTFs, $H_{\mathrm{em}}$ and $H_{\mathrm{ex}}$, and object spectrum $\Gamma$. The aberration-corrected 2PSIM (AO-2PSIM) image obtained from the object spectrum $\Gamma$ recovered by the dual-deconvolution algorithm is shown in Fig. 2c. The originally blurred image was made sharper, with a resolution approaching $\lambda_{\mathrm{em}}/4$. The estimated $H_{\mathrm{em}}$ and $H_{\mathrm{ex}}$ are visualized in Figs. 2d and 2e, respectively. The OTF bandwidths were substantially reduced, and the PTFs exhibited complex phase distributions due to aberrations. We obtained the excitation and emission PSFs by applying the Fourier transform to the respective OTF maps (Figs. 2f and 2g).

The performance of our aberration correction algorithm was quantified by comparing the intensity profiles and the MTFs of the reconstructed images in Figs. 2h and 2i, respectively. The red curve in Fig. 2i represents the MTF of the AO-2PSIM image in Fig. 2c, which closely matched the ideal MTF obtained from an aberration-free 2PSIM image (green curve). In contrast, the MTFs of the 2PFM and 2PSIM images without aberration correction, shown as gray and blue curves, respectively, exhibited a substantial loss of information at high frequencies.

To emphasize the advantages of our methodology, which corrects both excitation and emission PSFs, we compared it to a conventional single-PSF-based blind deconvolution algorithm (Supplementary Fig. S2). The conventional blind deconvolution marginally improved both 2PFM and 2PSIM images, and it was unable to achieve the resolution and contrast enhancement provided by our proposed method.

### Experimental Validation of Multiphoton Super-resolution Imaging

For the experimental validation of the proposed concept, we constructed a custom-made two-photon fluorescence imaging system equipped with a scientific CMOS camera at the detector plane (see Methods and Supplementary Information for details). A wavelength-tunable femtosecond pulsed laser (INSIGHT X3, Spectra Physics) served as the excitation source, and a high-numerical-aperture objective (N60X-NIR, Nikon, 60×, 1.0 NA) was used to focus the excitation. Image magnification at the camera was adjusted to ensure the camera's pixel pitch satisfied the Nyquist sampling interval for the emission cut-off frequency, $k_{\mathrm{em}}$. We captured two-photon fluorescence images by 2D-scanning the focused excitation beam. These images were used to construct the incoherent response matrix $f(\mathbf{r}_{\mathrm{d}};\mathbf{r}_{\mathrm{i}})$ and reconstruct 2PFM and AO-2PSIM images. Here, the scanning interval $r_{\mathrm{sc}}$ determines the maximum spatial frequency for excitation according to the Nyquist theorem, $k_{\mathrm{ex}} < 2\pi/(2r_{\mathrm{sc}})$. Specifically, the interval could be set to $r_{\mathrm{sc}} = \lambda_{\mathrm{ex}}/2\alpha n$ to achieve the diffraction-limited cut-off frequency for $n$-photon excitation.

To verify two-photon super-resolution imaging capability, we used 100-nm-diameter polystyrene beads labeled with Rhodamine B (see Methods for sample preparation). The excitation and peak emission wavelengths were $\lambda_{\mathrm{ex}} = 850$ nm and $\lambda_{\mathrm{em}} = 567$ nm, respectively, yielding a theoretical bandwidth-limited resolution of ~121 nm. The scanning interval was set to $r_{\mathrm{sc}} = 131$ nm. The reconstructed 2PFM, 2PSIM, and AO-2PSIM images are shown in Figs. 3a-c, respectively, with corresponding zoom-in regions shown in Fig. 3d. The reconstructed 2PFM and 2PSIM images are the results of conventional blind deconvolution, providing a fair comparison with the AO-2PSIM image. Line profiles along the dashed lines in Fig. 3d are displayed in Fig 3e. Compared to 2PFM, the 2PSIM image exhibited a resolution enhancement and optical sectioning due to aperture synthesis. However, AO-2PSIM demonstrated a substantial resolution improvement, clearly resolving the two beads with a separation of 240 nm, which is comparable to the theoretical diffraction-limited resolution of 2PFM, $\lambda_{\mathrm{ex}}/4 = 212.5$ nm. Figure 3f presents the spatial frequency spectra of the reconstructed images, clearly showing the extended spectral bandwidth of AO-2PSIM. Corresponding radially averaged spectra are displayed in Fig. 3g.

The excitation and emission OTFs estimated by the dual deconvolution algorithm are shown in Fig. 3h. While PTFs exhibited a relatively flat profile due to the absence of sample-induced aberrations, MTFs demonstrated a decrease at high spatial frequencies. This effect was more pronounced in the emission OTF due to its increased susceptibility to system aberrations caused by its shorter wavelength. By correcting both PTFs and MTFs, our algorithm enabled recovery of the full OTF bandwidth, achieving a resolution close to the theoretical limit.

### Experimental Validation of Multiphoton AO-2PSIM for Severe Aberrations

We experimentally validated the performance of the proposed multiphoton super-resolution imaging for targets under an aberration layer that introduces substantial aberrations. The first sample, consisting of Alexa 488 stained gold particles with an average diameter of 100 nm, was placed under an artificial aberrating layer (see Methods for sample preparation). The excitation and peak emission wavelengths were $\lambda_{\mathrm{em}} = 900$ nm and $\lambda_{\mathrm{ex}} = 520$ nm, respectively, yielding a theoretical bandwidth-limited resolution of ~121 nm. The scanning interval was set to $r_{\mathrm{sc}} = 131$ nm. Figures 4a-4c show the 2PFM, 2PSIM, and AO-2PSIM images, respectively. The excitation and emission OTF maps, along with the corresponding PSFs, are presented in Figs. 4d and 4e, respectively. The estimated OTF maps encompass both the system aberrations and those induced by the sample. Both excitation and emission MTF maps fall sharply from the center, significantly reducing the effective MTF bandwidths of 2PFM and 2PSIM images. Both the excitation and emission PSFs recovered from the corresponding OTFs show pronounced distortion. In particular, the excitation PSF was split into two spots mainly because the PTF was modulated along the $k_x$ direction. Consequently, each particle appeared as two particles in 2PFM and 2PSIM images, as indicated by the arrowheads in Figs. 4a and 4b. However, our algorithm was able to correct this artifact and recover a single sharp PSF. Furthermore, we could normalize the sharp drop of MTFs at high spatial frequencies using the identified OTFs, leading to the recovery of spatial resolution. The resolution, estimated as the full width at half maximum (FWHM) of five distinct particles in the AO-2PSIM image, was approximately 174 nm, which fell short of the theoretical limit of 121 nm. We attribute this to imperfect recovery of the target's MTF at high spatial frequencies due to insufficient signal-to-noise ratio (SNR). Nevertheless, the achieved resolution is still surpassed the diffraction limit of two-photon imaging, $\lambda_{\mathrm{ex}}/4 = 225$ nm, confirming the ability of our AO-2PSIM to achieve super-resolution imaging even under severe aberration conditions.

We also validated the proposed method for even more pronounced aberrations. A fluorescent resolution target was fabricated by placing a thin metal mask of an etched USAF target pattern onto a Rhodamine B solution (Fig. 4f). A scattering layer was superimposed on top of the metal mask to introduce the aberration (see Methods for sample preparation). Figures 4g-4i show 2PFM, 2PSIM, and AO-2PSIM images. The AO-2PSIM image shows a clear, high-contrast structure, whereas the 2PFM and 2PSIM images are severely blurred with multiple ghost artifacts due to strong aberrations. The third smallest line pairs with a spacing of 250 nm at the resolution target were clearly distinguished in the AO-2PSIM image.

The estimated excitation and emission OTFs with corresponding PSFs are shown in Figs. 4j and 4k, respectively. Both the excitation and emission PSFs are severely blurred and exhibit multiple foci, indicating strong wavefront distortions arising from the scattering medium. These multi-focal PSFs are responsible for the ghost artifacts in the images. Our algorithm retrieves both PTFs and MTFs over high spatial-frequency components and computationally corrects aberration and MTF attenuation. Once again, our dual-deconvolution algorithm finds both the excitation and emission OTFs without need for any prior knowledge, thereby correcting both the excitation and emission aberrations. This is a clear advantage, especially in the presence of strong aberrations, in recovering the resolving power compared to conventional single-PSF-based blind deconvolution.

**Two-Photon Super-resolution Imaging in Cells and Tissues**

We demonstrate the super-resolution imaging capabilities of our AO-2PSIM method in imaging cells and thick biological tissues (see Methods for sample preparation). Figure 5a shows a conventional 2PFM image of microtubules stained with Alexa 488 within a fixed COS-7 cell, and Fig. 5b shows the corresponding AO-2PSIM image. Excitation and peak emission wavelengths were $\lambda_{\mathrm{ex}} = 900$ nm and $\lambda_{\mathrm{em}} = 520$ nm, respectively. The 2PFM image exhibits noise, and the microtubule structures appear blurry. Applying aberration correction by our dual deconvolution method leads to significant improvements in both resolution and contrast. This is evident from the left panel in Fig. 5c, which shows the line profiles along the red dotted lines in the insets of Figs. 5a and 5b. We estimated the resolution based on the average width of a single microtubule branch, indicated by the dotted yellow boxes in the insets. The corresponding line profiles are also shown in the right panel in Fig. 5c. The resolution of the AO-2PSIM image was measured to be 134 nm, whereas the resolution of the 2PFM image was 280 nm.

Next, we performed two-photon imaging of ex-vivo mouse brain tissues immunolabeled with Thy1-EGFP. The excitation and peak emission wavelengths were 900 nm and 510 nm, respectively, yielding a theoretical bandwidth-limited resolution of ~120 nm. Figures 5d and 5e display the reconstructed 2PFM and AO-2PSIM images, respectively, at a depth of 130 μm. The 2PFM image (Fig. 5d) reveals somewhat blurred dendritic structures, with the necks either indistinct or invisible. In contrast, the AO-2PSIM image (Fig. 5e) offers a clear view of dendrites and their associated spines. By measuring the widths of the spine necks (Fig. 5f), we confirmed that our AO-2PSIM method can achieve super-resolution exceeding the diffraction limit, even for tissue imaging. Figures 5g and 5h show the 2PFM and AO-2PSIM images at a depth of 180 μm. The increased depth results in a more blurred 2PFM image compared to that at a depth of 130 μm. However, the AO-2PSIM image maintains high resolution and SNR, clearly visualizing the dendritic spine heads and necks even at this depth. Although the narrowest neck width at this depth is slightly larger than the theoretical super-resolution limit, it remains smaller than the diffraction limit (Fig. 5i).

We also conducted two-photon imaging of Zebrafish hindbrain to demonstrate the capability of AO-2PSIM in handing spatially varying aberrations[17], where the isoplanatic patch size is relatively small (see Supplementary Information for more information about isoplanatic patch analysis).

## Discussion

In this study, we introduced a multiphoton super-resolution fluorescence imaging technique via dual deconvolution of the incoherent response matrix, which offers computational correction of complex sample-induced aberrations and achieves deep-tissue super-resolution imaging. The proposed method allows for the reconstruction of an object image with a spatial resolution twice the diffraction limit by expanding the spatial frequency bandwidth through aperture synthesis. Our method has advantages compared to hardware AO in that it does not require complex wavefront shaping devices or wavefront sensing systems. It simply replaces a PMT or photodiode in a conventional laser scanning microscope with an array detector, such as a CMOS camera. Moreover, the proposed method does not require guide stars or any prior knowledge about the sample, making it applicable to a wide range of samples. Additionally, while most hardware AO systems correct aberrations in either the excitation or emission path[11], our method corrects both paths, resulting in better resolution recovery.

The unique feature of the proposed dual-deconvolution algorithm lies in that it is a matrix decomposition. This enables independent identification and correction of excitation and emission PSFs without any prior knowledge of the PSFs. This is crucial for achieving the best possible resolution in super-resolution imaging when dealing with severe aberrations. Conventional blind-deconvolution algorithms, which identify a single effective PSF given by the product of excitation and emission PSFs from a single blurred image, is a vector decomposition. Therefore, they can be effective only for mild aberrations with known PSF shapes, and their performance in resolution recovery is intrinsically lower than our dual-deconvolution algorithm. Our framework is so general that it can work for any scanning microscopy relying on incoherent emissions. In our present study, we demonstrated two-photon imaging, a representative deep-tissue imaging modality, of dendritic spines in thick mouse brain tissues with a spatial resolution of 130 nm up to a depth of 180 μm.

Our proposed computational AO is simpler than hardware AO in its implementation, but it has drawbacks. High spatial frequency components of the OTF are attenuated by aberrations and often obscured by noise. Although our dual-deconvolution approach provides the best possible recovery of high-frequency content, it is difficult to recover the information lost at the time of matrix recording. Another drawback is the long acquisition time. Our method relies on using an array detector to acquire images, which is associated with image scanning microscopy. This results in slower image acquisition compared to conventional confocal or multiphoton microscopes. The primary factor influencing imaging speed is the camera exposure time needed to capture the fluorescence maps, typically requiring around 1 ms per each scanning point. Thus, acquiring a 100×100-pixel image takes about 10 s. However, employing multifocal illumination or speckle illumination techniques[37, 38] could significantly acquisition time, potentially enabling real-time imaging for most biological studies. After acquisition, the image processing time needed to construct and decompose the incoherent response matrix for a 100×100-pixel image is less than 1 s on a graphics processing unit (GPU, GeForce RTX 3090, NVIDIA).

Given the simplicity of the hardware and the significant improvements in resolution and imaging depth offered by our proposed image reconstruction algorithm, we anticipate rapid integration of our technique into commercial multiphoton microscopy systems. Future strategies include enhancing image acquisition speed through methods such as parallel imaging with multifocal illumination[39, 40] using a digital micromirror device or a liquid-crystal spatial light modulator[41-45]. Additionally, employing SNR-optimized array detectors such as single-photon avalanche diodes (SPAD) array detector[29] may facilitate capturing raw fluorescence maps at a higher SNR, thereby enabling the recovery of the full bandwidth of OTFs.

## Methods

### Matrix decomposition algorithm

In the matrix formalism, the spectral IRM $\boldsymbol{F}$ in Eq. (3) is given by the Hadamard product of the effective OTF matrix $\boldsymbol{H}$ and object spectrum matrix $\boldsymbol{\Gamma}$. The goal of the dual-deconvolution algorithm is to estimate $H_{\mathrm{em}}$, $H_{\mathrm{ex}}$, and $\Gamma$ which minimize the squared Frobenius norm of the difference of the measured spectral IRM $\boldsymbol{F}_{\mathrm{mea}}$ and the model: $\underset{\boldsymbol{H}_{\mathrm{em}},\boldsymbol{\Gamma},\boldsymbol{H}_{\mathrm{ex}}}{\text{mininize}} \|\boldsymbol{F}_{\mathrm{mea}} - \boldsymbol{H} \circ \boldsymbol{\Gamma}\|_{\mathrm{F}}{}^{\mathbf{2}}$, subject to $\Gamma(\mathbf{k}_{\mathrm{o}}, \mathbf{k}_{\mathrm{i}}) = \Gamma(\boldsymbol{\Delta}\mathbf{k} = \mathbf{k}_{\mathrm{o}} - \mathbf{k}_{\mathrm{i}})$, and $H(\mathbf{k}_{\mathrm{o}}, \mathbf{k}_{\mathrm{i}}) = H_{\mathrm{em}}(\mathbf{k}_{\mathrm{o}}) H_{\mathrm{ex}}(\mathbf{k}_{\mathrm{i}})$.

The basic principle of our method is based on the iterative Weiner filter method[35, 36], but successively estimates three unknowns, $\Gamma$, $H_{\mathrm{em}}$, and $H_{\mathrm{em}}$. In each iteration, the algorithm sequentially updates $\Gamma$, $H_{\mathrm{em}}$, and $H_{\mathrm{ex}}$ in Wiener filter-like equations:

1) Estimate object spectrum $\Gamma(\mathbf{k}_{\mathrm{o}}, \mathbf{k}_{\mathrm{i}})$ by applying a matrix Wiener filter $W(\mathbf{k}_{\mathrm{o}}, \mathbf{k}_{\mathrm{i}}) = \frac{H^*(\mathbf{k}_{\mathrm{o}}, \mathbf{k}_{\mathrm{i}})}{|H(\mathbf{k}_{\mathrm{o}}, \mathbf{k}_{\mathrm{i}})|^2 + \alpha^2}$ to $F(\mathbf{k}_{\mathrm{o}}, \mathbf{k}_{\mathrm{i}}) = H(\mathbf{k}_{\mathrm{o}}, \mathbf{k}_{\mathrm{i}}) \Gamma(\mathbf{k}_{\mathrm{o}}, \mathbf{k}_{\mathrm{i}})$, followed by $\Gamma(\boldsymbol{\Delta}\mathbf{k}) = \sum_{\mathbf{k}_{\mathrm{i}}} \Gamma(\mathbf{k}_{\mathrm{o}}, \mathbf{k}_{\mathrm{i}})$.
2) Update the emission OTF $H_{\mathrm{em}}(\mathbf{k}_{\mathrm{o}})$ from the correlation between each row of $F(\mathbf{k}_{\mathrm{o}}, \mathbf{k}_{\mathrm{i}})$ and $G(\mathbf{k}_{\mathrm{o}}, \mathbf{k}_{\mathrm{i}}) = \Gamma(\mathbf{k}_{\mathrm{o}}, \mathbf{k}_{\mathrm{i}}) H_{\mathrm{ex}}(\mathbf{k}_{\mathrm{i}})$.
3) Estimate the object spectrum $\Gamma(\mathbf{k}_{\mathrm{o}}, \mathbf{k}_{\mathrm{I}})$ again with the newly estimated $H_{\mathrm{em}}(\mathbf{k}_{\mathrm{o}})$.
4) Update the excitation OTF $H_{\mathrm{ex}}(\mathbf{k}_{\mathrm{i}})$ from the correlation between each column of $F(\mathbf{k}_{\mathrm{o}}, \mathbf{k}_{\mathrm{i}})$ and $G(\mathbf{k}_{\mathrm{o}}, \mathbf{k}_{\mathrm{i}}) = H_{\mathrm{em}}(\mathbf{k}_{\mathrm{o}}) \Gamma(\mathbf{k}_{\mathrm{o}}, \mathbf{k}_{\mathrm{i}})$.

The Algorithm for dual deconvolution is given in the following:

---

**Algorithm:** Dual Deconvolution

---

1: **input:** spectral IRM $F(\mathbf{k}_\mathrm{o}, \mathbf{k}_\mathrm{i})$

2: **initialize:** $H_\mathrm{em}(\mathbf{k}_\mathrm{o}) = 1,\ H_\mathrm{ex}(\mathbf{k}_\mathrm{i}) = 1$

3: **for** $n = 0,1,...,$ until stopping criterion is reached

4: **update** $\Gamma$**:**

5: $W(\mathbf{k}_\mathrm{o}, \mathbf{k}_\mathrm{i}) = \frac{H^*(\mathbf{k}_\mathrm{o},\mathbf{k}_\mathrm{i})}{|H(\mathbf{k}_\mathrm{o},\mathbf{k}_\mathrm{i})|^2+\alpha^2}$, where $H(\mathbf{k}_\mathrm{o}, \mathbf{k}_\mathrm{i}) = H_\mathrm{em}(\mathbf{k}_\mathrm{o})H_\mathrm{ex}(\mathbf{k}_\mathrm{i})$

6: $\Gamma(\Delta\mathbf{k}) = \frac{N(\Delta\mathbf{k})}{|N(\Delta\mathbf{k})|^2+\beta^2}\sum_{\mathbf{k}_\mathrm{i}} W(\mathbf{k}_\mathrm{i} + \mathbf{\Delta k}, \mathbf{k}_\mathrm{i})F(\mathbf{k}_\mathrm{i} + \mathbf{\Delta k}, \mathbf{k}_\mathrm{i})$,

7: where $N(\Delta\mathbf{k}) = \sum_{\mathbf{k}_\mathrm{i}} H(\mathbf{k}_\mathrm{i} + \mathbf{\Delta k}, \mathbf{k}_\mathrm{i})W(\mathbf{k}_\mathrm{i} + \mathbf{\Delta k}, \mathbf{k}_\mathrm{i})$

8: **update** $H_\mathrm{em}$:

9: $H_\mathrm{em}(\mathbf{k}_\mathrm{o}) \leftarrow \frac{\sum_{\mathbf{k}_\mathrm{i}\neq\mathbf{0}}|\Gamma(\mathbf{k}_\mathrm{o},\mathbf{k}_\mathrm{i})H_\mathrm{ex}(\mathbf{k}_\mathrm{i})|^2}{\left(\sum_{\mathbf{k}_\mathrm{i}\neq\mathbf{0}}|\Gamma(\mathbf{k}_\mathrm{o},\mathbf{k}_\mathrm{i})H_\mathrm{ex}(\mathbf{k}_\mathrm{i})|^2\right)^2+\gamma^2}\sum_{\mathbf{k}_\mathrm{i}\neq\mathbf{0}} F(\mathbf{k}_\mathrm{o}, \mathbf{k}_\mathrm{i})\Gamma^*(\mathbf{k}_\mathrm{o}, \mathbf{k}_\mathrm{i})H^*_\mathrm{ex}(\mathbf{k}_\mathrm{i})$

10: **update** $\Gamma$: repeat line 5-7

11: **update** $H_\mathrm{ex}$:

12: $H_\mathrm{ex}(\mathbf{k}_\mathrm{i}) \leftarrow \frac{\sum_{\mathbf{k}_\mathrm{o}\neq\mathbf{0}}|\Gamma(\mathbf{k}_\mathrm{o},\mathbf{k}_\mathrm{i})H_\mathrm{em}(\mathbf{k}_\mathrm{o})|^2}{\left(\sum_{\mathbf{k}_\mathrm{o}\neq\mathbf{0}}|\Gamma(\mathbf{k}_\mathrm{o},\mathbf{k}_\mathrm{i})H_\mathrm{em}(\mathbf{k}_\mathrm{o})|^2\right)^2+\gamma^2}\sum_{\mathbf{k}_\mathrm{o}\neq\mathbf{0}} F(\mathbf{k}_\mathrm{o}, \mathbf{k}_\mathrm{i})\Gamma^*(\mathbf{k}_\mathrm{o}, \mathbf{k}_\mathrm{i})H^*_\mathrm{em}(\mathbf{k}_\mathrm{o})$

13: **end for**

For Wiener filters, damping factors $\alpha$, $\beta$, and $\gamma$ are used as small real numbers proportional to the signal-to-noise ratios. The iteration stops when the difference of the effective OTFs between successive iterates converge to a predetermined threshold. Note that when updating $H_\mathrm{em}(\mathbf{k}_\mathrm{o})$ and $H_\mathrm{ex}(\mathbf{k}_\mathrm{i})$ at lines 9 and 12 in the algorithm, the summations over the spatial frequencies exclude the zero-frequency term. For objects with a significant DC spectral component, such as speckle-like objects, accurate decomposition of the excitation OTF and emission OTF can be challenging, leading to slower algorithm convergence.

**Numerical generation of an incoherent response matrix in Fig. 2**

The simulation employed an emission wavelength of 520 nm, a numerical aperture of $\alpha = 1$, and a theoretical bandwidth-limited resolution of 130 nm, which is a quarter of the emission wavelength. The full set number of scans was 242 × 242 positions with a scan interval of 130 nm, which covers the region of interest (ROI) of 31.5 × 31.5 μm$^2$. The excitation and emission pupil maps were computationally prepared by the random superposition of Zernike modes numbering up to 30. These pupil aberrations were autocorrelated to derive OTFs, which were subsequently converted into intensity PSFs by fast-Fourier transform. We simulated the point-illumination and wide-field detection by convolving the target with the two independently generated excitation and emission PSFs. Each fluorescence image is sampled at the region of detection (ROD) of $100 \times 100$ pixels with a pixel resolution of 130 nm. Therefore, the full basis points in our simulation resulted in a raw data set of $(100 \times 100) \times (242 \times 242)$ pixels corresponding to the ROI of $242 \times 242$ pixels and ROD of $100 \times 100$ pixels. The IRM is constructed, containing $(242 \times 242) \times (242 \times 242)$ pixels after resizing ROD to ROI. The spectral IRM was obtained by the Fourier transform of IRM with respect to column and row.

**Experimental setup**

We built a two-photon microscope with a wavelength-tunable pulsed laser (INSIGHT X3, Spectra Physics). The excitation wavelength was chosen depending on the types of fluorophores. For example, an excitation wavelength of 900 nm was utilized for Alexa 488, GFP, and eGFP, and 850 nm was used for Rhodamine B. A short-pass dichroic mirror (DMSP680B, Thorlabs) was positioned to separate the emitted fluorescence from the excitation laser beam. The excitation laser beam underwent raster-scanning through 2D Galvano mirrors (GVS002, Thorlabs) before being focused onto the sample plane via a high-numerical-aperture objective (N60X-NIR, Nikon, ×60, 1.0 NA). Fluorescence emissions were captured by the same objective lens, de-scanned by the 2D Galvano mirrors, and subsequently traversed through the previous short-pass dichroic mirror, a short-pass filter (FESH0700, Thorlabs), and a band-pass filter to eliminate the stray excitation laser beam. The band-pass filter was chosen according to the peak emission wavelengths of fluorophores. The peak emission wavelength was 530 nm for Alexa 488, GFP, and eGFP, and 565 nm for Rhodamine B. The filtered fluorescence signals were recorded by a sCMOS camera (pco.edge 4.2, PCO AG) in the de-scanned frame. A photomultiplier tube (PMT, H13543-20, Hamamatsu) was also installed for fast imaging to identify and localize fluorescence targets and fluorophores. The total acquisition time ranged from 52 to 260 seconds for an ROI of 30 × 30 μm with an exposure time of 1 to 5 ms, depending on the degree of aberrations.

**Preparation of test samples**

**Rhodamine B polystyrene beads and Alexa 488 gold nanoparticles sample preparation:** From 10% l-lysine diluted with distilled water, 1 ml of l-lysine was carefully dispensed onto a slide glass (P000BMBS, Marienfeld Superior), which was subsequently covered with a cover glass (P000BMCU, Marienfeld Superior) to suppress the droplet for a duration of 30 seconds. This procedure ensured the even distribution of l-lysine across the entire surface of the slide glass. Following a 30-minute drying period, the l-lysine- coated slide glass was rinsed with distilled water to remove any residual impurities. Meanwhile, a solution containing 100 μl of either Rhodamine B polystyrene beads (PS100-RB-1, Nanocs) or Alexa 488 gold nanoparticles (GFL-100, CD Bioparticles) was prepared with 5~10 ml of distilled water in a disposable beaker. Subsequently, an appropriate volume of this nanoparticle mixture was applied to the slide glass, dependent on the particle concentration within the defined region of interest. The fluorescence particles underwent a chemical attachment process with the l-lysine on the slide glass surface. After allowing 10 minutes for the mixture to dry, the slide was rinsed with distilled water and left to air-dry in preparation for the subsequent particle imaging demonstration.

**Artificial scattering medium preparation:** A pure hardened PDMS layer of 150 μm thickness was placed onto a slide glass, positioned fluorescence dyed beads or particles. A procedure for aberration scattering, a cleaning polymer solution (FCDFR, First Contact) was subsequently applied over the PDMS layer. The surface of the cleaning polymer was intentionally roughened by scratching it with a sandpaper, creating different structures of slow-varying grating patterns to fast-varying randomized patterns. These surface curvatures that create wavefront error are randomly generated as the polymer solution gradually dried in room temperature conditions.

**Fluorescence USAF resolution target fabrication:** First, we deposited a 40 nm thickness of titanium metal and a 40 nm thickness of gold metal layers on a cover glass with a thickness of 500 μm. The titanium metal serves as an adhesion layer between gold and glass, and both gold and titanium metal act as a metal mask to block the excitation and emission. Next, we created a customized USAF resolution target pattern using the FIB etcher. Through an etching process that removed the entire 80 nm of gold and titanium, the residual metal on the target pattern formed a mask to conceal background fluorescence.

Following this, we prepared a solution of Rhodamine B (MFCD00011931, Sigma-Aldrich) in ethanol, diluted to one-hundredth of the maximum solubility of Rhodamine B in ethanol. The fluorescence solution was then poured into a well created by a double-sided sticker spacer (654002, Grace Bio-Labs). The fabricated target mask was flipped over and stuck on the spacer. Finally, a custom-made scattering medium with a cleaning polymer solution (FCDFR, First Contact) was attached to the cover glass surface. With the metal mask positioned between the scattering medium and the fluorescence solution, the fabricated sample served as a well-defined 2D extended fluorescence target, as illustrated in Fig. 4f.

**Preparation of biological samples:**

**COS-7 cell preparation:** In the COS-7 cell imaging experiment, AC28806 cells from the Korean Collection for Type Cultures (KCTC) were cultured on cover glasses. The cells were cultured overnight in DMEM (Gibco) with 10% FBS (Thermo Fisher) and 1% penicillin-streptomycin (Thermo Fisher). Before fixation, the cells were rinsed with pre-warmed PBS(Corning) and treated with pre-warmed extraction buffer (37°C) containing 0.125% Triton X-100 (Sigma-Aldrich) and 0.4% glutaraldehyde (Sigma-Aldrich) in PBS. After rinsing three times with PBS, the cells were fixed using pre-warmed fixation buffer (37°C) containing 3.2% paraformaldehyde (Biosesang) and 0.1% glutaraldehyde for 10 minutes at room temperature (RT). Following three PBS rinses, the cells were permeabilized using a solution containing 3% BSA (Sigma Life Science) and 0.5% Triton X-100 in PBS. Subsequently, the cells were incubated with a primary antibody targeting tubulin (ab6046, Abcam), diluted 1000-fold in a blocking buffer (3% BSA and 0.5% Triton X-100 in PBS) for 1 hour at RT. After primary antibody incubation, the cells were rinsed with PBS and exposed to a secondary antibody labeled with Alexa Fluor 488 (A-11006, Thermo Fisher). The secondary antibody, also diluted 1000-fold in the blocking buffer, was applied to the cells for 1 hour at RT with agitation. The cells underwent an additional three PBS rinses and were then stored at 4°C.

**Mouse brain preparation:** In the mouse brain imaging study, 12 weeks Thy1-EGFP lime M mice (Jackson Labs #007788) were anesthetized with an intraperitoneal injection of 100 mg/kg ketamine and 10 mg/kg xylazine before decapitation. Their brains were promptly extracted and placed in an ice-cold artificial cerebrospinal fluid (ACSF). The brains were then sliced into 400~500 μm-thick coronal sections with a vibroslicer (World Precision Instruments, USA) and fixed at 4°C in 4% paraformaldehyde overnight. For imaging, the fixed brain was washed with PBS three times and then stuck to plastic dish and immersed in PBS. The entire experimental process was carried out with the approval of the Committee of Animal Research Policy at Korea University (approval number KUIACUC-2022-0013).

## Acknowledgments

This work was supported by the Institute for Basic Science (IBS-R023-D1) and the National Research Foundation of Korea (NRF) grant funded by the Korea government (MSIT) (No. RS-2023-00213310, No. RS-2024-00442818).

## Data availability

The data that support the findings of this study are available from the authors on reasonable request.

## Code availability

The code used in this study are available from the corresponding author upon reasonable request.

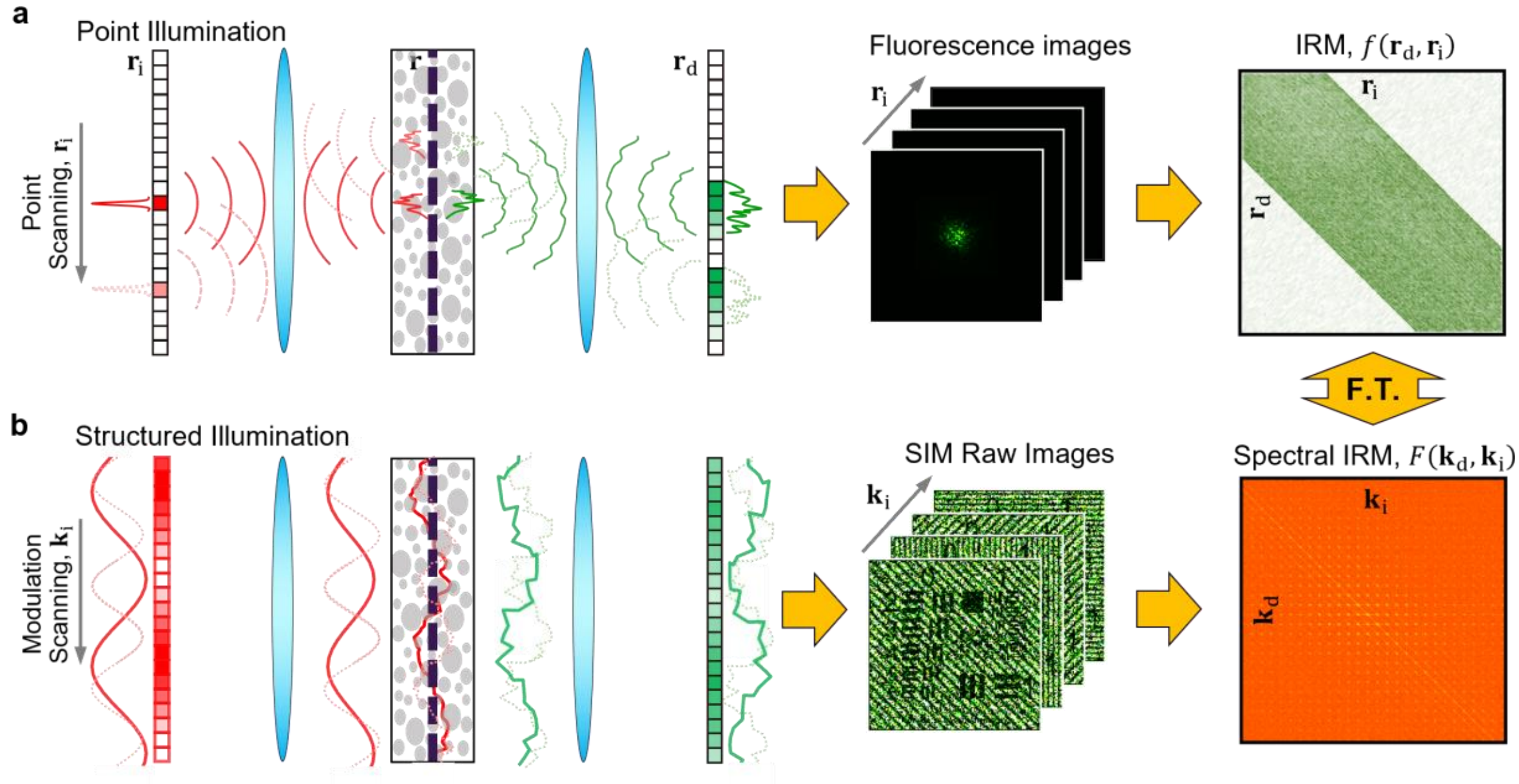

**Fig. 1 | Incoherent response matrix (IRM) in the space and spatial frequency domains. a,** Workflow of IRM measurement: A tightly focused excitation beam is scanned along the coordinate $\mathbf{r}_i$ conjugate to the sample plane $\mathbf{r}$, where fluorophore target is embedded inside a scattering medium. A series of wide-field fluorescence images for point illuminations are recorded by an array detector with detector coordinate $\mathbf{r}_d$. Both the excitation and emission PSFs are distorted due to sample-induced aberration and scattering. IRM, $f(\mathbf{r}_d, \mathbf{r}_i)$, is constructed from the recorded fluorescence images. **b,** Construction of spectral IRM. Sinusoidal intensity pattern, $I_{ill}(\mathbf{r}_i) = e^{i\mathbf{k}_i \cdot \mathbf{r}_i}$, with modulation frequency $\mathbf{k}_i$ is considered as the illumination, and the resulting fluorescence image $I_{fl}(\mathbf{r}_d, \mathbf{k}_i)$, is synthesized from the recorded IRM in (**a**). The spectral IRM, $F(\mathbf{k}_d, \mathbf{k}_i)$, can be constructed by applying a Fourier transform to $I_{fl}(\mathbf{r}_d, \mathbf{k}_i)$ with respect to $\mathbf{r}_d$. Alternatively, $F(\mathbf{k}_d, \mathbf{k}_i)$ can be obtained directly by Fourier transforming $f(\mathbf{r}_d; \mathbf{r}_i)$ in (**a**) with respect to both $\mathbf{r}_i$ and $\mathbf{r}_d$.

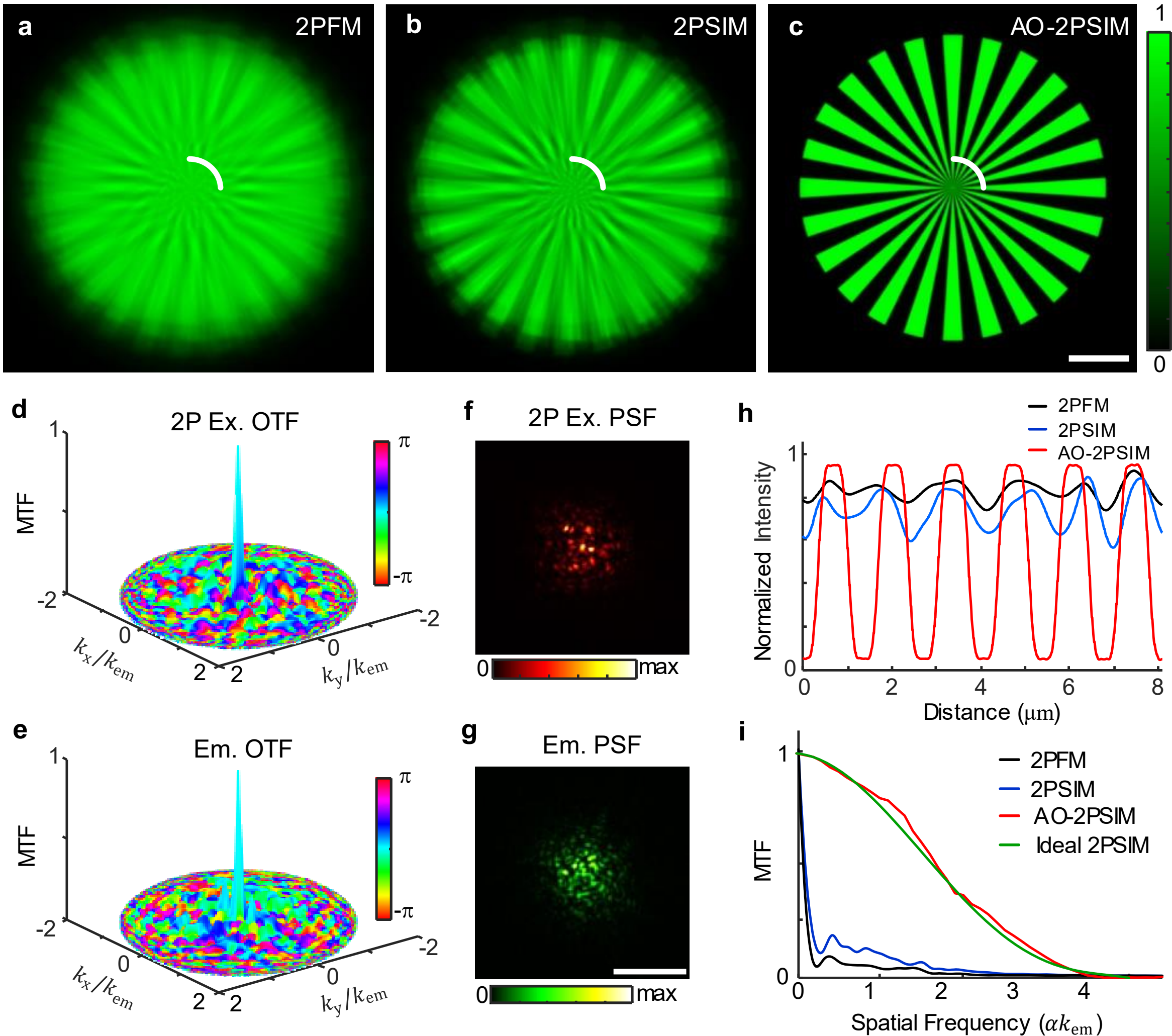

**Fig. 2 | Aberration correction by dual deconvolution of simulated target. a,** 2PFM image reconstructed from a simulated IRM. **b,** 2PSIM image obtained from the aperture synthesis of the spectral IRM. **c,** AO-2PSIM image recovered by the dual deconvolution algorithm. **d-e,** Estimated excitation and emission OTF maps, respectively. The height represents the MTF, while the color represents the PTF. The spectral frequencies, $k_x$ and $k_y$, are in units of $k_{\text{em}}$. **f-g,** Excitation and emission PSFs obtained by the inverse Fourier transform of the estimated excitation and emission OTFs, respectively. **h,** Intensity profiles along the white curves in (**a-c**). **i,** MTFs obtained from the 2PFM, 2PSIM, and AO-2PSIM images in (**a-c**). MTF values for each spatial frequency were determined from the contrast values of the intensity profiles in (**h**), with the radius corresponding to the spatial frequency. The ideal 2PSIM represents a 2PSIM image obtained in the absence of aberrations. Scale bars in (**c**, **g**), 5 μm.

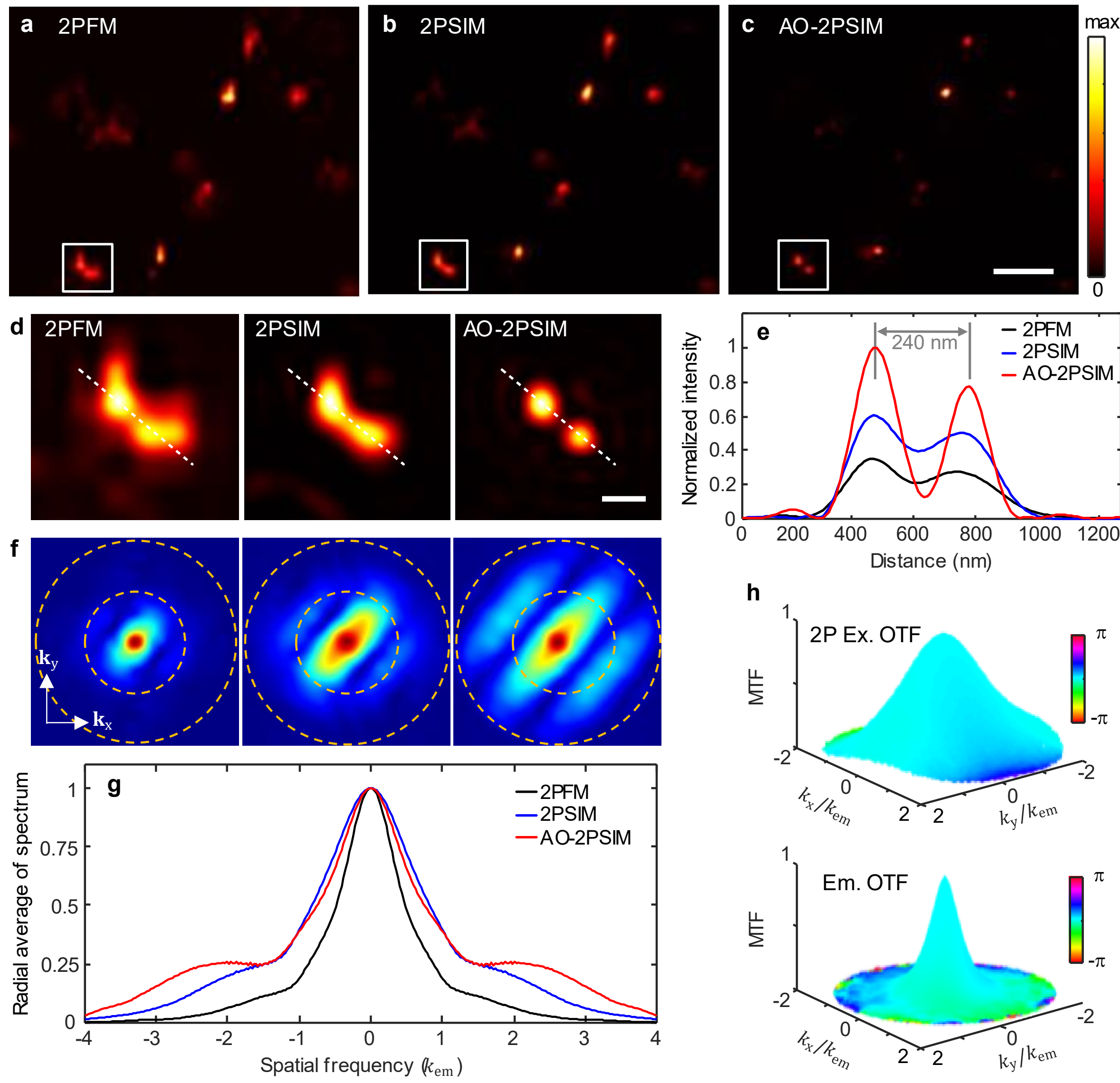

**Fig. 3 | Two-photon super-resolution imaging of 100-nm-diameter Rhodamine B polystyrene beads. a-c,** 2PFM, 2PSIM, and AO-SIM images. The reconstructed 2PFM and 2PSIM images are the results of conventional blind deconvolution. **d,** Zoomed-in regions of the white squares in (**a-c**), showing 2PFM (left), SIM (middle), and AO-SIM (right) images. **e,** Line profiles along the dashed lines in (**d**). The AO-2PSIM image clearly resolves two peaks with a measured distance of 240 nm. **f,** Spatial frequency spectra of 2PFM, 2PSIM, and AO-2PSIM images in (**d**). Two dashed circles indicate bandwidths with radii of $2k_{em}$ and $4k_{em}$. AO-2PSIM exhibits the largest frequency bandwidth, close to full bandwidth of $4k_{em}$. **g,** Radial average of spectra are measured from (**f**), corresponding to the extended spectral bandwidth of AO-2PSIM. **h,** Excitation OTF (top) and emission OTF (bottom), visualized in 4D. Height represents MTF and colormap represents PTF. Scale bars in (**a-c**) and (**d**), 1 μm and 200 nm, respectively.

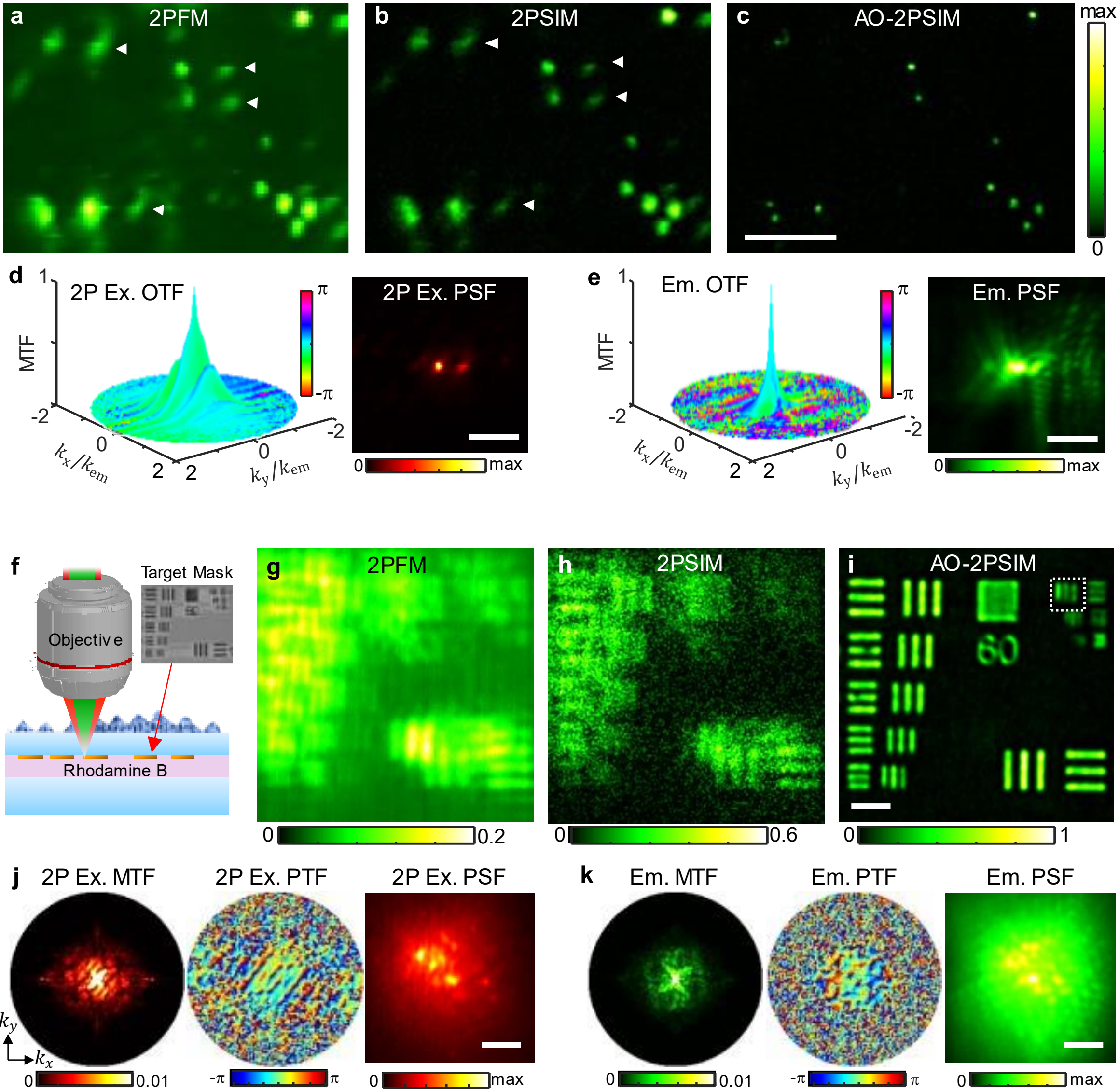

**Fig. 4 | AO-2PSIM under aberrating layers. a-c,** 2PFM, SIM, and AO-SIM images of 100-nm-diameter Alexa-488-stained gold particles under an artificial aberrating medium. The 2PFM and 2PSIM images are the results of conventional blind deconvolution. **d,** Excitation OTF (left) and the corresponding PSF (right), identified by the dual deconvolution algorithm. **e,** Identified emission OTF (left) and the corresponding PSF (right). **f,** Schematic of imaging a custom-made fluorescent resolution target. The resolution target is placed beneath a scattering medium, introducing substantial aberrations. The target mask is fabricated on a coverglass, and a Rhodamine B solution is placed underneath the resolution target. **g-i,** 2PFM, 2PSIM, and AO-2PSIM images of the resolution target, respectively. The 2PFM and 2PSIM images are the results of conventional blind deconvolution. The third smallest line spacing of the target (red dotted box in (**i**)) is 250 nm. **j,** Excitation MTF (left), PTF (middle), and PSF (right), respectively, identified by the dual deconvolution. **k,** Emission MTF (left), PTF (middle), and PSF (right), respectively. Scale bars in the figures, 3 μm.

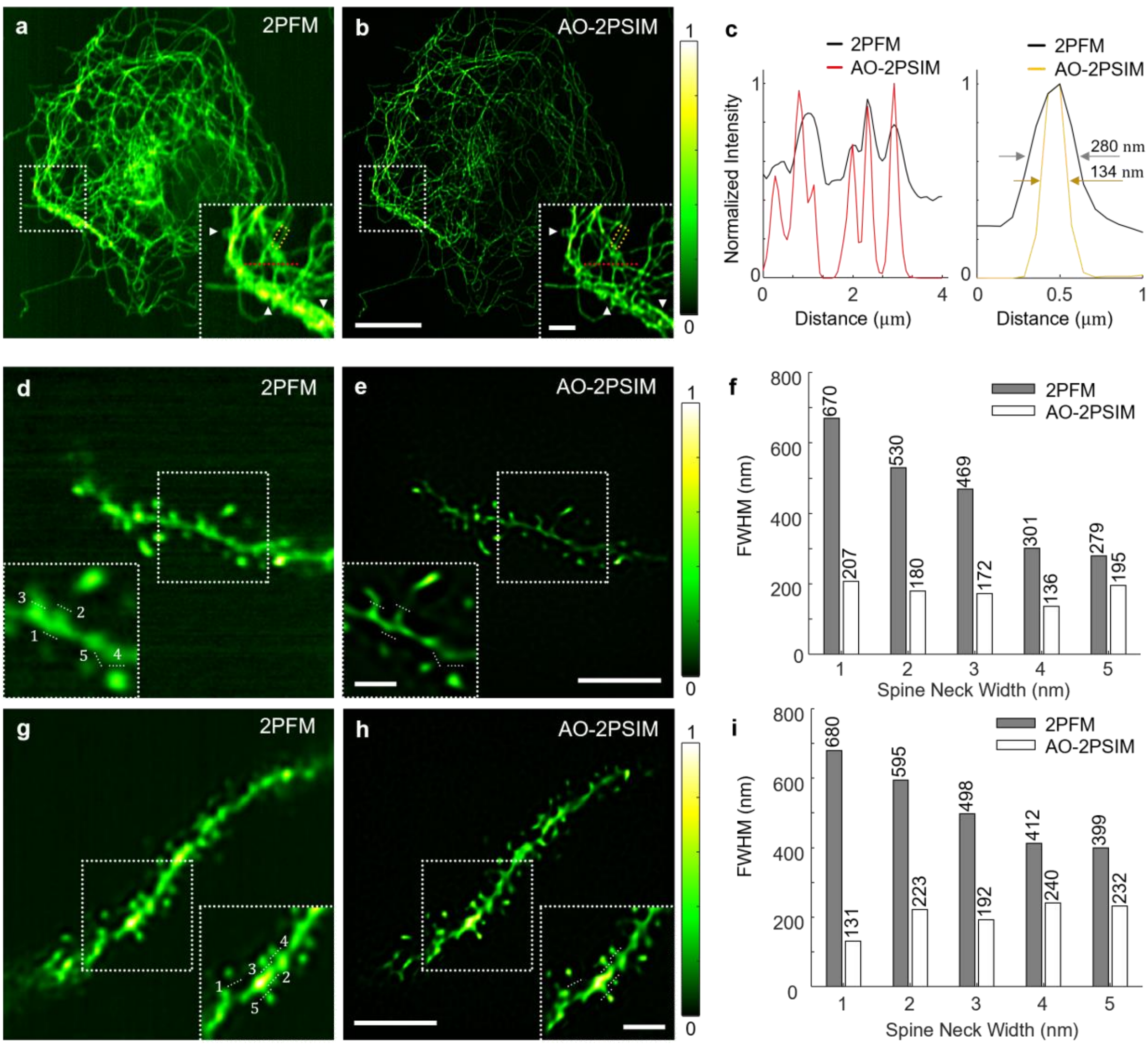

**Fig. 5 | AO-2PSIM imaging of biological samples. a-b,** Imaging of microtubules inside a whole COS-7 cell, stained with Alexa-488 fluorescence conjugate. 2PFM and AO-2PSIM images are shown in (**a**) and (**b**), respectively. The insets on the bottom right of each figure are the zoomed-in images of dashed rectangular boxes. Complex tubular structures depicted by white arrowheads are well visualized in the AO-2PSIM image. **c,** Line profiles along the red dotted lines in the insets are shown on the left panel. The average intensity profiles of a single branch of microtubule in the yellow dotted boxes are shown on the right panel. **d-e,** Imaging of Thy1-EGFP immunolabeled ex-vivo mouse brain tissue. 2PFM and AO-2PSIM images of dendritic spines at a depth of 130 μm are shown in (**d**) and (**e**), respectively. Zoomed-in images of dashed rectangular boxes are shown at the bottom left of each figure. **f,** The measured widths of the spine necks labeled as 1-5 in the insets are shown. **g-i,** Same as (**d-f**), but taken at a depth of 180 μm. Scale bars: Cell images (**a, b**), 10 μm; Ihe insets of (**a**, **b**), 2 μm; Dendrite images (**e, h**), 5 μm; Insets of (**e, h**); 2 μm.